\documentclass[paper]{JHEP} 
\usepackage{epsfig,multicol,cite}
\def \lsim
{\raisebox{-3pt}{$\>\stackrel{<}{\scriptstyle\sim}\>$}}

\title{Soft-Gluon Resummation\\
for  Bottom  Fragmentation\\
in Top Quark Decay}
\author{Matteo Cacciari\\
Dipartimento di Fisica, Universit\`a di Parma, Italy,\\
and INFN, Sezione di Milano, Gruppo Collegato di Parma\@.\\
E-mail: \email{Matteo.Cacciari@cern.ch}}
\author{Gennaro Corcella\\
Max-Planck-Institut f\"ur Physik, Werner-Heisenberg-Institut,\\
F\"ohringer Ring 6, D-80805 M\"unchen, Germany\@.\\
E-mail: \email{corcella@mppmu.mpg.de}}
\author{Alexander D. Mitov\thanks{The work of A.D.M. was supported in part 
by the U.S. Department of Energy, under grant DE-FG02-91ER40685.}\\
Department of Physics and Astronomy, University of Rochester,\\
Rochester, NY 14627, U.S.A.\\
E-mail: \email{amitov@pas.rochester.edu}}

\preprint{
UPRF-2002-12\\
MPI-PhT 2002-34}

\abstract{
We study soft-gluon radiation in top quark decay within the framework of
perturbative fragmentation functions. We present results
for the $b$-quark energy distribution, accounting for
soft-gluon
resummation to next-to-leading logarithmic accuracy in both the
$\overline{\mathrm{MS}}$
coefficient function and in the initial condition of
the perturbative fragmentation function.
The results show a remarkable improvement and the $b$-quark energy spectrum
in top quark decay exhibits very little dependence on factorization
and renormalization scales.
We present some hadron-level results in both $x_B$ and moment space by
including non-perturbative information determined from $e^+e^-$ data.}

\begin{document} 

\section{Introduction}
Heavy-flavour and in particular top quark physics is presently one of the main
fields of investigation in theoretical and experimental particle physics.
The current experiments at the Tevatron accelerator and, ultimately,
at LHC \cite{lhc}
and $e^+e^-$ Linear Colliders \cite{lc} will produce large amounts of
top quark pairs, which will allow one to perform improved measurements of
top properties, such as its mass.
For this purpose, precise calculations for top production and decay
processes are mandatory.

While fixed-order calculations reliably predict total cross sections
or widths, differential distributions typically contain large logarithms
associated with soft or collinear parton radiation.
For heavy-quark production processes,
although the quark mass $m$ (much larger than the QCD scale
$\Lambda$) acts as a regulator for the collinear
singularity, event shapes still contain large  $\alpha_S^n\ln^p(Q^2/m^2)$
(with $p\le n$) terms,
$Q$ being
a typical scale of the process, which make fixed-order predictions
unreliable when $Q\gg m$.
Such logarithms can be resummed by using the
perturbative fragmentation approach\cite{mele}, which factorizes the rate of
heavy-quark production into the convolution of a
coefficient function, describing the emission of a massless parton,
and a perturbative fragmentation function
$D(\mu_F,m$), where $\mu_F$ is the factorization scale.
The perturbative fragmentation function expresses the
transition of the massless parton into the massive quark, and its value
at any scale $\mu_F$ can be obtained by solving the
Dokshitzer--Gribov--Lipatov--Altarelli--Parisi
(DGLAP) evolution
equations \cite{ap,dgl} once an initial condition at a scale
$\mu_{0F}$ is given. In ref.~\cite{mele} the large $\ln(Q^2/m^2)$
collinear logarithms were resummed to next-to-leading logarithmic (NLL)
accuracy in the case of heavy quark production in $e^+e^-$ collisions
and an explicit next-to-leading order (NLO) expression for $D(\mu_F,m$),
which was argued to be
process independent, was given. More recently, ref.~\cite{cc} has
established the process independence in a more general way.

The approach of perturbative fragmentation has been extensively used for
$e^+e^-$ annihilation \cite{cc,colnas,cagre1,nasole,canas},
hadron collisions \cite{cagre2,cagre3}, photoproduction
\cite{cagre4,cagre1}
and, more recently, for bottom quark production in top quark decay $t\to bW$
\cite{cormit}.

The initial condition of the perturbative fragmentation
function and the coefficient function, though free of the collinear
large $\ln(Q^2/m^2)$, contain large logarithms which are
due to soft-gluon radiation. The ones contained in the initial
condition are process independent~\cite{cc}, and were already included
in the top-to-bottom decay process in~\cite{cormit}, where the complete
${\cal O}(\alpha_S)$ calculation of the $t\to bW(g)$ process was also
performed.
The large
logarithms contained in the coefficient function are instead process
dependent, and have to be evaluated for every specific case.
It is precisely the purpose of this paper to
extend the analysis of ref.~\cite{cormit} and to
present results for soft-gluon resummation in the coefficient
function. This will allow, together with the results of~\cite{cc}, to
complete the evaluation of soft-gluon effects to NLL accuracy for the
top-to-bottom decay process, and to investigate the impact on the $b$-quark
energy distribution.

After the fragmentation of heavy quarks in $e^+e^-$ collisions
considered in~\cite{cc},  this is the first process whose large
logarithms (both collinear and soft) are fully  resummed to NLL
accuracy within the perturbative fragmentation function formalism. The
consistency of this perturbative description with the one used in the
$e^+e^-$ process makes it possible to fit non-perturbative information
from $e^+e^-$ data and use it to make predictions.
We shall therefore be able
to predict the spectrum for  $b$-flavoured hadron energy distributions
in top decay using $e^+e^-$ experimental data from LEP.

The outline of the paper is as follows.
In section 2 we review bottom quark production in top quark decay
within the framework of perturbative fragmentation. In section 3 we
present analytic results for the NLL soft-gluon resummation of the
coefficient function in top decay. At the end of the section we also
comment on the relation between our results and previous work on
soft-gluon resummation in heavy-flavour
decay~\cite{korster,roth,aglietti,aglric}. In
section 4 we show the $b$-quark energy distribution in top decay and
investigate the impact of soft-gluon resummation. In section 5 we
discuss inclusion of non-perturbative effects and present results
for $b$-flavoured hadron spectra in top decay.  In section 6 we
summarize our main results.

\section{Perturbative fragmentation and top quark decay}
We consider top decay into a bottom quark and a real $W$ boson plus,
to order $\alpha_S$, a gluon:
\begin{equation}
t(p_t)\to b(p_b)W(p_W)\left( g(p_g) \right)
\label{dec}
\end{equation}
\noindent
and define the bottom and gluon normalized energy fractions $x_b$ and
$x_g$:
\begin{equation}
x_b={1\over{1-w}}{{2 p_b\cdot p_t}\over {m_t^2}},\ \ \
x_g={1\over{1-w}}{{2 p_g\cdot p_t}\over {m_t^2}},
\label{xbpart}
\end{equation}
where $w=m_W^2/m_t^2$.
Neglecting the $b$ mass, we have $0\leq x_{b,g} \leq 1$.

Order $\alpha_S$ corrections to the decay process (\ref{dec}) were
considered in~\cite{cormit}. It was observed there that, since $m_b \ll m_t$,
one can readily neglect $m_b/m_t$ power suppressed terms, but on
the other hand it is
important to resum to all orders terms enhanced, at order
$\alpha_S$, by the
presence of $\ln(m_t^2/m_b^2)$. Such a resummation was performed
in~\cite{cormit} by employing the perturbative fragmentation
formalism~\cite{mele,cc}:
The differential width for the production of a
massive $b$ quark in top decay is written in terms of the
the convolution
\begin{eqnarray}
{1\over {\Gamma_B}} {{d\Gamma}\over{dx_b}} (x_b,m_t,m_W,m_b) &=&
\sum_i\int_{x_b}^1
{{{dz}\over z}\left[{1\over{\Gamma_B}}
{{d\hat\Gamma_i}\over {dz}}(z,m_t,m_W,\mu_F)
\right]^{\overline{\mathrm{MS}}}
D_i\left({x_b\over z},\mu_F,m_b \right)} \nonumber \\
&&+ {\cal
O}\left((m_b/m_t)^p\right) \; ,
\label{pff}
\end{eqnarray}
\noindent
where $\Gamma_B$ is the width of the Born process
$t\to bW$, $d\hat\Gamma_i /dz$ is the differential width for the production of
a massless parton $i$ in top decay with energy fraction $z$,
$D_i(x,\mu_F,m_b)$ is the perturbative
fragmentation function for a parton $i$ to fragment
into a massive $b$ quark,
$\mu_F$ is the  factorization scale. The term ${\cal
O}\left((m_b/m_t)^p\right)$ on the right-hand side stands for
contributions that are suppressed by some power $p$ ($p\ge 1$) of $m_b$ in
the $m_b \ll m_t$ regime. Of course, non-perturbative corrections of the
type $\Lambda/m_t$ and $\Lambda/m_b$ are understood on the right-hand
side of eq.~(\ref{pff}). We
shall use everywhere a branching fraction $B(t\to bW)=1$, and only
include $i=b$ in the above summation.

The massless differential distribution $(1/\Gamma_B )~d\hat\Gamma_i /dz$
(which is
what we shall also refer to as ``coefficient function'') is defined in
the  $\overline{\mathrm{MS}}$
factorization scheme after subtraction of the collinear singularities.
It has been calculated at order $\alpha_S$ in~\cite{cormit}. In the
following we shall often use its Mellin moments, defined by
\begin{equation}
\hat\Gamma_N  =
\int_0^1 {dz  \ z^{N-1}
{1\over{\Gamma_B}}{{d\hat\Gamma}\over{dz}}(z) }.
\end{equation}
In moment space the convolution (\ref{pff}) can then be rewritten as
\begin{equation}
\Gamma_N(m_t,m_W,m_b) = \hat\Gamma_N(m_t,m_W,\mu_F) D_{b,N}(\mu_F,m_b)
\,.
\label{gamman}
\end{equation}

The perturbative fragmentation function $D_b(x,\mu_F,m_b)$
at any scale $\mu_F$ can be obtained by solving the DGLAP equations.
As shown in \cite{cc}, as long as one can neglect contributions
proportional to powers of $(m_b/m_t)^p$,
the initial condition for the perturbative
fragmentation function, which we evaluate at a scale $\mu_{0F}$,
is process independent (but scheme dependent). In the
$\overline{\mathrm{MS}}$ scheme it reads~\cite{mele}:
\begin{equation}
D_b^{\rm ini}(x,\mu_{0F},m_b)=\delta(1-x)+{{\alpha_S(\mu_0^2)C_F}\over{2\pi}}
\left[{{1+x^2}\over{1-x}}\left(\ln {{\mu_{0F}^2}\over{m_b^2}}-
2\ln (1-x)-1\right)\right]_+.
\label{dbb}
\end{equation}
The solution of the DGLAP equations in the non-singlet sector,
for the evolution from the scale
$\mu_{0F}$ to $\mu_F$, is given in moment space by:
\begin{eqnarray}
D_{b,N}(\mu_F,m_b)&=&
D_{b,N}^{\rm ini}(\mu_{0F},m_b)\exp\left\{ {{P_N^{(0)}}\over{2\pi b_0}}
\ln{{\alpha_S(\mu^2_{0F})}\over {\alpha_S(\mu^2_F)}}\right.\nonumber\\
&+&
\left.{{\alpha_S(\mu^2_{0F})-\alpha_S(\mu^2_F)}\over{4\pi^2b_0}}
\left[P_N^{(1)}-{{2\pi b_1}\over {b_0}}P_N^{(0)}\right]\right\},
\label{dresum}
\end{eqnarray}
In eq.~(\ref{dresum})
$P_N^{(0)}$ and $P_N^{(1)}$ are the Mellin transforms of the leading and
next-to-leading order Altarelli-Parisi splitting vertices, and their explicit
expression can be found, e.g., in \cite{mele}.
$b_0$ and $b_1$ are the first two coefficients of the QCD $\beta$-function
\begin{equation}
b_0={{33-2n_f}\over {12\pi}},\ \ b_1={{153-19n_f}\over{24\pi^2}},
\end{equation}
which enter the following expression for the strong coupling constant at
a scale $Q^2$:
\begin{equation}
\alpha_S(Q^2)={1\over {b_0\ln(Q^2/\Lambda^2)}}
\left\{ 1-{{b_1\ln\left[\ln (Q^2/\Lambda^2)\right]}\over
{b_0^2\ln(Q^2/\Lambda^2)}}\right\}.
\label{alpha}
\end{equation}
Equation~(\ref{dresum}) resums to all order terms containing large
$\ln(\mu_F^2/\mu_{0F}^2)$.
In particular, leading
($\alpha_S^n\ln^n(\mu_F^2/\mu_{0F}^2)$)
and next-to-leading ($\alpha_S^{n}
\ln^{n-1}(\mu_F^2/\mu_{0F}^2)$)  logarithms are resummed.
Setting, as done in \cite{cormit},
$\mu_F\simeq m_t$ and $\mu_{0F}\simeq m_b$, one resums 
the large $\ln(m_t^2/m_b^2)$ terms with NLL accuracy.
These are indeed the large collinear logarithms exhibited by the fixed-order
calculation with  a massive $b$ quark \cite{cormit}.

\section{Soft-gluon resummation}
In this section we address the problem of soft-gluon resummation in top quark
decay.
The $\overline{\mathrm{MS}}$ coefficient function computed in \cite{cormit}
and the initial condition of the perturbative fragmentation function
(\ref{dbb}) contain terms behaving like
$1/(1-x)_+$ or $[\ln(1-x)/(1-x)]_+$,
which become arbitrarily large when $x$ approaches one.
This is equivalent to contributions proportional to
$\ln N$ and $\ln^2 N$ in moment space, as can be seen
by writing the $\overline{\mathrm{MS}}$ coefficient
function~\cite{cormit} in the
large-$N$ limit\footnote{Following \cite{aglric}, we note that, by
defining $n = N \exp(\gamma_E)$, we could rewrite this expression in terms
of $\ln(n)$ rather than $\ln N$, with no $\gamma_E$ terms explicitly
appearing.}:
\begin{eqnarray}
\hat\Gamma_N (m_t,m_W,\mu_F)
&= & 1+{{\alpha_S C_F}\over{2\pi}}
\left\{2\ln^2 N +\left[4\gamma_E +2-4\ln(1-w)
-2\ln{{m_t^2}\over{\mu_F^2}}\right]\ln N\right.\nonumber\\
&+&K(m_t,m_W,\mu_F)+{\cal O}\left( {1\over N}\right)\Bigg\}
\label{largen}
\end{eqnarray}
where $\gamma_E = 0.577...$ is the Euler constant and $w =
m_W^2/m_t^2$, as defined in section 1.
In eq.~(\ref{largen}) we have introduced the function
$K(m_t,m_W,\mu_F)$,
which contains terms which are constant with respect to $N$. It reads:
\begin{eqnarray}
K(m_t,m_W,\mu_F)&=&
\left({3\over 2}-2\gamma_E\right)\ln{{m_t^2}\over{\mu_F^2}}+2\gamma_E^2
+2\gamma_E\left[1-2\ln(1-w)\right]\nonumber\\
&+& 2\ln w\ln(1-w)-2{{1-w}\over{1+2w}}\ln(1-w)-
{{2w}\over{1-w}}\ln w\nonumber\\
&+&4 {\mathrm{Li}}_2(1-w) -6-{{\pi^2}\over 3}\, .
\label{kappa}
\end{eqnarray}

The $x\to 1$ ($N\to\infty $) limit corresponds to soft-gluon
radiation in top decay. These soft logarithms need to be resummed to all
orders in $\alpha_S$ \cite{ct,sterman} to improve our prediction.

Soft logarithms in the initial condition of the perturbative
fragmentation function are process independent. We can hence 
resum them with NLL accuracy using the result presented in \cite{cc},
which we do not report here for the sake of brevity.
We present instead the  results for the NLL resummation of process-dependent
soft-gluon contributions in the
$\overline{\mathrm{MS}}$ coefficient function.

In order to resum the large terms in eq.~(\ref{largen}),
we follow standard techniques \cite{ct}, 
evaluate the amplitude of the process in eq.~(\ref{dec})
at ${\cal O}(\alpha_S)$
in the eikonal approximation and exponentiate the result.
The eikonal current reads:
\begin{equation}
|J(p_t,p_b,p_g)|^2=\left| {{m_t^2}\over{(p_t\cdot p_g)^2}}-
2{{(p_t\cdot p_b)}\over{(p_t\cdot p_g)(p_b\cdot p_g)}}\right|.
\end{equation}
For the sake of comparison with \cite{ct},
we express the ${\cal O}(\alpha_S)$ width in the 
soft approximation as an integral
over the variables\footnote{We point out that our definition of the
integration variable
$q^2$ is analogous to the quantity $(1-z)k^2$ to which the authors of
ref.~\cite{ct} set the scale for $\alpha_S$ for soft-gluon
resummation in Drell--Yan and
Deep-Inelastic-Scattering processes. For small-angle
radiation, $q^2\simeq q_T^2$, the gluon transverse momentum with respect to
the $b$-quark line. The variable $z$ is analogous 
to $z=1-E_g/E_q$ of ref.~\cite{ct}.}
$q^2=(p_b+p_g)^2x_g$ and $z=1-x_g$, with
$0\leq q^2\leq m_t^2(1-w)^2(1-z)^2$ and $0\leq z\leq 1$.
The limits $z\to 1$ and $q^2\to 0$ correspond to soft and collinear emission
respectively. In soft approximation, $z\simeq x_b$, the $b$-quark energy 
fraction. We obtain: 
\begin{eqnarray}
\hat\Gamma_N(m_t,m_W,\mu_F)&=&{{C_F}\over{\pi}}
\int_0^1{dz {{z^{N-1}-1}\over {1-z}}}
\left[\int_{\mu_F^2}^{m_t^2(1-w)^2(1-z)^2}{{dq^2}\over{q^2}}
\alpha_S\right.\nonumber\\
&-&\left.{1\over{m_t^2(1-w)^2(1-z)^2}}
\int_0^{m_t^2(1-w)^2(1-z)^2}{dq^2 \alpha_S}\right].
\label{eik}
\end{eqnarray}
In eq.~(\ref{eik}) we have regularized the collinear
singularity setting the cutoff $q^2\geq \mu_F^2$. 
At NLL accuracy level, this is equivalent to
$\overline{\mathrm{MS}}$ subtraction in dimensional regularization
\cite{cc,cmw}.

In order to perform soft-gluon resummation to NLL accuracy a number of
operations have to be performed on this expression.
We set the argument of $\alpha_S$ equal to $q^2$ and, as far as the 
collinear-divergent term is concerned, we perform the replacement
\begin{equation}
{{C_F}\over\pi}{{\alpha_S(q^2)}\over{q^2}}
\to {{A\left[\alpha_S(q^2)\right]}\over {q^2}},
\label{afun}
\end{equation}
where the function $A(\alpha_S)$ was introduced in \cite{ct} and is
detailed below.
Moreover, the integral over $q^2$ of the non-collinear divergent term
can be written, up to terms beyond NLL accuracy, as 
\begin{equation}
{1\over{m_t^2(1-w)^2(1-z)^2}}
\int_0^{m_t^2(1-w)^2(1-z)^2}{dq^2\alpha_S(q^2)}=
\alpha_S\left(m_t^2(1-w)^2(1-z)^2\right).
\label{asint}
\end{equation}
This term describes soft radiation at large-angle, 
i.e. not collinear enhanced, and it 
is characteristic of processes where a heavy quark (the top quark in our case, the bottom quark in
\cite{korster,roth,aglietti,aglric}) is present.
It can be generalized to all orders by replacing eq.~(\ref{asint}) according 
to:
\begin{equation}
-{{C_F}\over{\pi}}\alpha_S\left(m_t^2(1-w)^2(1-z)^2\right)\to
S\left[\alpha_S\left(m_t^2(1-w)^2(1-z)^2\right)\right].
\label{hfun}
\end{equation}
This function is called
$\Gamma(\alpha_S)$ in \cite{korster}, $B(\alpha_S)$ in
\cite{roth}, $S(\alpha_S)$ in \cite{aglietti}, $D(\alpha_S)$ in
\cite{aglric}.
We now insert Eqs.~(\ref{afun}-\ref{hfun}) into eq.~(\ref{eik})
and exponentiate the result. We obtain:
\begin{eqnarray}
\label{resum}
\ln \Delta_N &=&
\int_0^1 {dz {{z^{N-1}-1}\over{1-z}}}
\left\{\int_{\mu_F^2}^{m_t^2(1-w)^2(1-z)^2}
{{dq^2}\over {q^2}} A\left[\alpha_S(q^2)\right]\right. \nonumber\\
&+&
S\left[\alpha_S\left(m_t^2(1-w)^2(1-z)^2\right)\right]\Bigg\}.
\label{deltares}
\end{eqnarray}
We would like to evaluate eq.~(\ref{deltares}) to NLL level.
The function $A(\alpha_S)$ can be expanded
as follows:
\begin{equation}
A(\alpha_S)=\sum_{n=1}^{\infty}\left({{\alpha_S}\over {\pi}}\right)^n
A^{(n)}.
\end{equation}
The first two coefficients are needed at NLL level and are given
by~\cite{ct,kt}:
\begin{equation}
A^{(1)}=C_F,
\end{equation}
\begin{equation}
A^{(2)}=
{1\over 2} C_F \left[ C_A\left( {{67}\over{18}}-{{\pi^2}\over 6}\right)
-{5\over 9}n_f\right],
\end{equation}
where $C_F = 4/3$, $C_A = 3$ and $n_f$ is the number of quark flavours,
which we shall take equal to five for bottom production.

The function $S(\alpha_S)$ can be expanded according to:
\begin{equation}
S(\alpha_S)=\sum_{n=1}^{\infty}\left({{\alpha_S}\over {\pi}}\right)^n
S^{(n)}.
\end{equation}
At NLL level, we are just interested in the first term of the above
expansion, which is given by:
\begin{equation}
S^{(1)}=-C_F.
\end{equation}

The integral in eq.~(\ref{deltares}) can be performed, up to NLL
accuracy, by making the following
replacement \cite{ct}:
\begin{equation}
z^{N-1}-1\to -\Theta\left( 1-{{e^{-\gamma_E}}\over N}-z\right) \; ,
\end{equation}
$\Theta$ being the Heaviside step function.
This leads to writing the following result for the function
$\Delta_N$:
\begin{equation}
\Delta_N(m_t,m_W,\alpha_S(\mu^2),\mu,\mu_F)=\exp\left[\ln N g^{(1)}(\lambda)+
g^{(2)}(\lambda,\mu,\mu_F)\right]\, ,
\label{deltaint}
\end{equation}
with
\begin{equation}
\lambda=b_0\alpha_S(\mu^2)\ln N \, ,
\end{equation}
and the functions $g^{(1)}$ and $g^{(2)}$ given by
\begin{eqnarray}
g^{(1)}(\lambda)&=&
\frac{A^{(1)}}{2\pi b_0 \lambda} \;
[ 2\lambda + (1-2\lambda) \ln (1-2\lambda)] \;,\\
g^{(2)}(\lambda,\mu,\mu_F) &=&
\frac{A^{(1)}}{2 \pi b_0}\left[\ln \frac{m_t^2(1-w)^2}{\mu_F^2}
- 2\gamma_E\right] \ln(1-2\lambda)\nonumber\\
&+&\frac{A^{(1)}  b_1}{4 \pi b_0^3}
\left[ 4\lambda + 2 \ln (1-2\lambda) +
\ln^2 (1-2\lambda) \right]\nonumber\\
&-& \frac{1}{2\pi b_0} \left[2\lambda + \ln (1-2\lambda) \right]
\left(\frac{A^{(2)}}{\pi b_0} + A^{(1)}\ln\frac{\mu^2}{\mu_{F}^2}\right)
\nonumber\\
&+& \frac{S^{(1)}}{2\pi b_0} \ln (1-2\lambda).
\end{eqnarray}
In eq.~(\ref{deltaint}) the term $\ln N g^{(1)}(\lambda)$ accounts
for the resummation of leading logarithms $\alpha_S^n\ln^{n+1}N$
in the Sudakov exponent, while the function $g^{(2)}(\lambda,\mu,\mu_F)$
resums NLL terms $\alpha_S^n\ln^nN$.

Furthermore, we follow ref.~\cite{cc} and
in our final Sudakov-resummed coefficient
function we also include the constant terms of eq.~(\ref{kappa}):
 \begin{eqnarray}
\hat\Gamma_N^S(m_t,m_W,\alpha_S(\mu^2),\mu,\mu_F)&=&
\left[1+{{\alpha_S(\mu^2)\ C_F}\over{2\pi}}
K(m_t,m_W,\mu_F)\right]\nonumber\\
&\times &
\exp\left[\ln N g^{(1)}(\lambda)+g^{(2)}(\lambda,\mu,\mu_F)\right].
\label{delta}
\end{eqnarray}
One can check that
the ${\cal O} (\alpha_S)$ expansion of eq.(\ref{delta}) yields
eq.~(\ref{largen}).

We now match the resummed coefficient function to the exact
first-order result, so that also $1/N$ suppressed terms, which are
important in the region $x_b < 1$, are taken into account.
We adopt the same matching prescription as in \cite{cc}:
we add the resummed result to the exact coefficient function and,
in order to avoid double counting, we subtract what they have in common,
i.e. the up-to-${\cal O}(\alpha_S)$ terms in the expansion of
eq.~(\ref{delta}).
Our final result for the resummed coefficient function
reads\footnote{Alternative ways of matching, identical up to order
$\alpha_S$ and differing in higher-order, subleading terms, are of
course possible.}:
\begin{eqnarray}
\hat\Gamma_N^{\mathrm{res}}(m_t,m_W,\alpha_S(\mu^2),\mu,\mu_F)
&=&\hat\Gamma_N^S(m_t,m_W,\alpha_S(\mu^2),\mu,\mu_F)\nonumber\\
&-&
\left[\hat\Gamma_N^S(m_t,m_W,\alpha_S(\mu^2),\mu,\mu_F)
\right]_{\alpha_S}\nonumber\\
&+&\left[\hat\Gamma_N(m_t,m_W,\alpha_S(\mu^2),\mu,\mu_F)\right]_{\alpha_S},
\end{eqnarray}
where $[\hat\Gamma_N^S]_{\alpha_S}$ and
$[\hat\Gamma_N]_{\alpha_S}$ are respectively the expansion of
eq.~(\ref{delta}) up to ${\cal O}(\alpha_S)$
 and the  full fixed-order
top-decay coefficient function at ${\cal O} (\alpha_S)$,
evaluated in Appendix B of ref.~\cite{cormit}.

Before closing this section we would like to add more comments on the
comparison of our resummed expression
with other similar results obtained in heavy quark decay processes
\cite{korster,roth,aglietti,aglric}.

Besides the obvious replacement of a bottom quark with a top in the initial
state, our work presents other essential differences.
We have resummed large collinear logarithms $\alpha_S\ln(m_t^2/m_b^2)$,
while Refs.~\cite{korster,roth,aglietti,aglric} just address the decay of heavy
quarks into massless quarks.
Moreover, this work still differs in a critical issue. Those papers
are concerned with observing the lepton produced by the $W$
decay or the photon in the $b\to X_s \gamma$ process, while we wish instead
to observe  the outgoing $b$ quark.
This is immediately clear from the
choice of the $x$ variable whose $x\to 1$ endpoint leads to the Sudakov
logarithms. In our case it is the normalized
energy fraction of the outgoing bottom quark;
in \cite{korster,roth,aglietti,aglric} it is instead related
to the energy of either the lepton or the radiated photon.

The most evident effect of this different perspective is that an
additional scale, namely the invariant mass of the recoiling hadronic
jet, enters the results \cite{korster,roth,aglietti,aglric}, but it is
instead absent
in our case. An additional function (called $\gamma(\alpha_S)$ in
\cite{korster,roth}, $C(\alpha_S)$ in \cite{aglietti}, $B(\alpha_S)$ in
\cite{aglric}) appears in those papers.  The argument of $\alpha_S$ in
this function is related to the invariant mass of the unobserved final
state jet constituted by the outgoing quark and the gluon(s). It is
worth noting that an identical function, called
$B\left[\alpha_S\left(Q^2(1-z)\right)\right]$, 
also appears in the $e^+e^-$~\cite{cc} and
DIS~\cite{ct} massless
coefficient functions, where it is again
associated with
the invariant mass of the unobserved jet. 
We do not have instead
any $B(\alpha_S)$ contribution in our result.
In fact, this function  contains collinear radiation associated with an
undetected quark, which we do not have since the $b$ quark is observed.

We also observe that to order $\alpha_s$ the coefficient $S^{(1)}$ 
coincides with the corresponding
$H^{(1)}$ of  
the function $H\left[\alpha_S\left(m_b^2(1-z)^2\right)\right]$, 
which resums soft terms in the initial 
condition of the perturbative fragmentation function \cite{cc}. 
It will be very interesting to compare the functions $S(\alpha_S)$ and
$H(\alpha_S)$ at higher orders as well.

One final comment we wish to make is that, as expected, in our final result,
eq.(\ref{gamman}), which accounts for NLL soft resummation in
both the coefficient function and the initial condition of the perturbative
fragmentation function,
$\alpha_S^n\ln^{n+1} N$ terms do not appear, since they are due to soft
{\sl and} collinear radiation. Both the quarks being heavy,
only the former leads to a logarithmic enhancement.
Double logarithms are generated by a
mismatch in the lower and upper $q^2$ integration limit over the
$A[\alpha_S(q^2)]$ function in the exponent of the resummation
expression. In our case both of them have the same functional
dependence with respect to $z$, i.e. $(1-z)^2$ (see eq.~(\ref{resum}) and
eq.~(69) of ref.~\cite{cc}).
The cancellation of the $\alpha_S^n\ln^{n+1} N$
term  can be explicitly
seen at order $\alpha_S$ by comparing the large-$N$ limit for the
coefficient function, eq.~(\ref{largen}), and the initial condition
(eq.~(45) of ref.~\cite{cc}): the $\ln^2 N$ terms have identical coefficients
and opposite signs.

\section{Energy spectrum of the $b$ quark}
In this section we  present results for the $b$-quark energy distribution
in top decay.
The $b$-quark spectrum in $N$-space $\Gamma_N(m_t,m_W,\mu_F)$ is given by
eq.~(\ref{gamman}).
In the following we shall normalize $\Gamma_N$ to the full NLO
width $\Gamma$ \cite{nlo}, so that $\Gamma_1 = 1$ will always hold.
Results in $x_b$-space will be obtained by inverting numerically
eq.~(\ref{gamman}) via contour integration in the complex plane,
using the minimal prescription \cite{cmnt} to avoid the Landau pole.

In order to estimate the effect of the NLL soft-gluon resummation, we
compare our result with ref.~\cite{cormit} and use the same values for the
parameters:
$m_t=175$~GeV, $m_b=5$~GeV, $m_W=80$~GeV and $\Lambda^{(5)}=200$~MeV.

In figure~\ref{fig1} we present the $x_b$ distribution according to the 
approach
of perturbative fragmentation, with and without NLL soft-gluon resummation.
For the scales appearing in Eqs.~(\ref{dbb}), (\ref{dresum})
and (\ref{largen}) we have set
$\mu_F=\mu=m_t$ and $\mu_0=\mu_{0F}=m_b$.
We note that the two distributions agree for $x_b\lsim 0.8$, while for
larger $x_b$ values the resummation of large terms $x_b\to 1$
smoothens out the distribution, which exhibits the Sudakov peak.
Both distributions become negative for $x_b\to 0$ and $x_b\to 1$.
As discussed in \cite{cormit}, the negative behaviour at small $x_b$
can be related to the presence of unresummed
$\alpha_S\ln x_b$ terms in the coefficient function.
At large $x_b$, we approach instead the non-perturbative region, and
resumming
leading and next-to-leading logarithms is still not sufficient to
correctly describe the spectrum for $x_b$ close to $1$.
In fact, the range of reliability of the perturbative calculation has been
estimated to be $x_b\lsim 1-\Lambda/m_b\simeq 0.95$ \cite{cc}.

\FIGURE[ht]{\epsfig{file=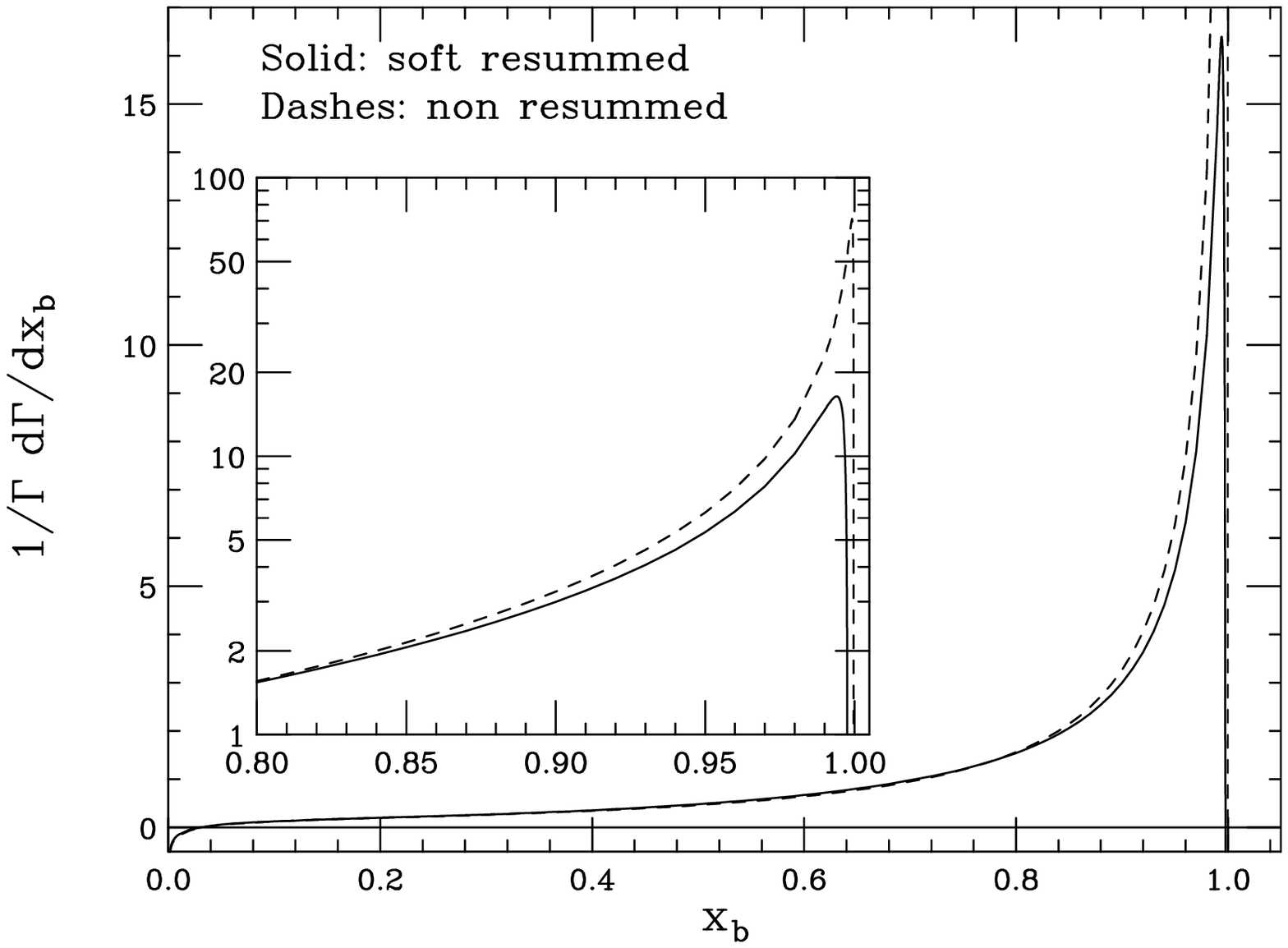,width=11.cm} 
        \caption{\small
$b$-quark energy distribution in top decay according to
the perturbative fragmentation approach, with (solid line) and without
(dashes) NLL soft-gluon resummation.
In the inset figure, we show the same curves on a logarithmic scale,
for $x_b>0.8$.
We have set $\mu_F=\mu=m_t$ and $\mu_{0F}=\mu_0=m_b$.}
\label{fig1}}
\FIGURE{\epsfig{file=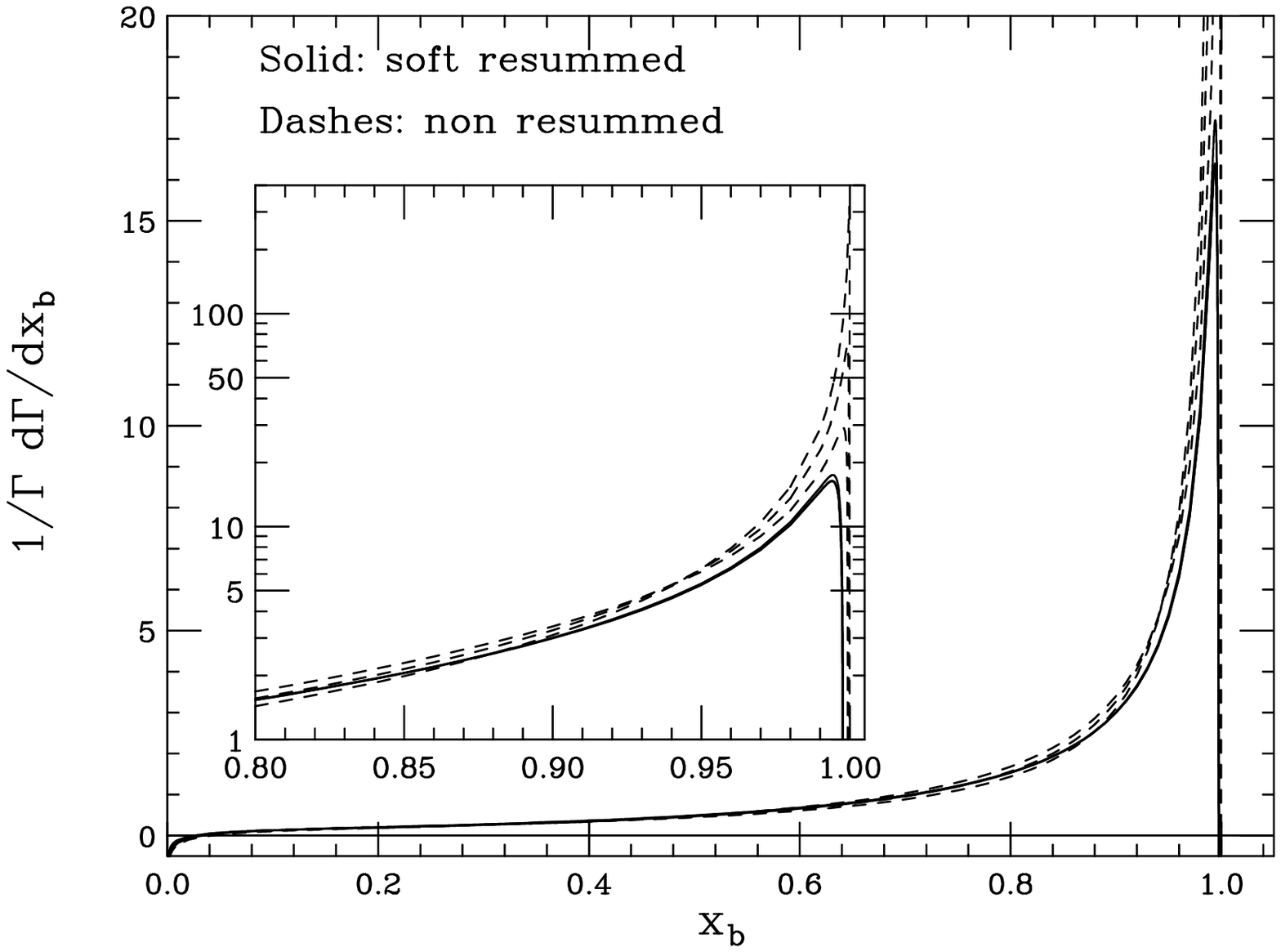,width=11.cm} 
\caption{\small $b$-quark energy spectrum for different values
of the factorization scale $\mu_F$, with (solid) and
without (dashes)
NLL soft-gluon resummation. The other scales are fixed
at $\mu=m_t$, $\mu_0=\mu_{0F}=m_b$.
As in figure~\ref{fig1}, 
in the inset figure, we present the same plots for large values
of $x_b$, on a logarithmic scale.}
\label{figfac}}
\FIGURE[ht]{\epsfig{file=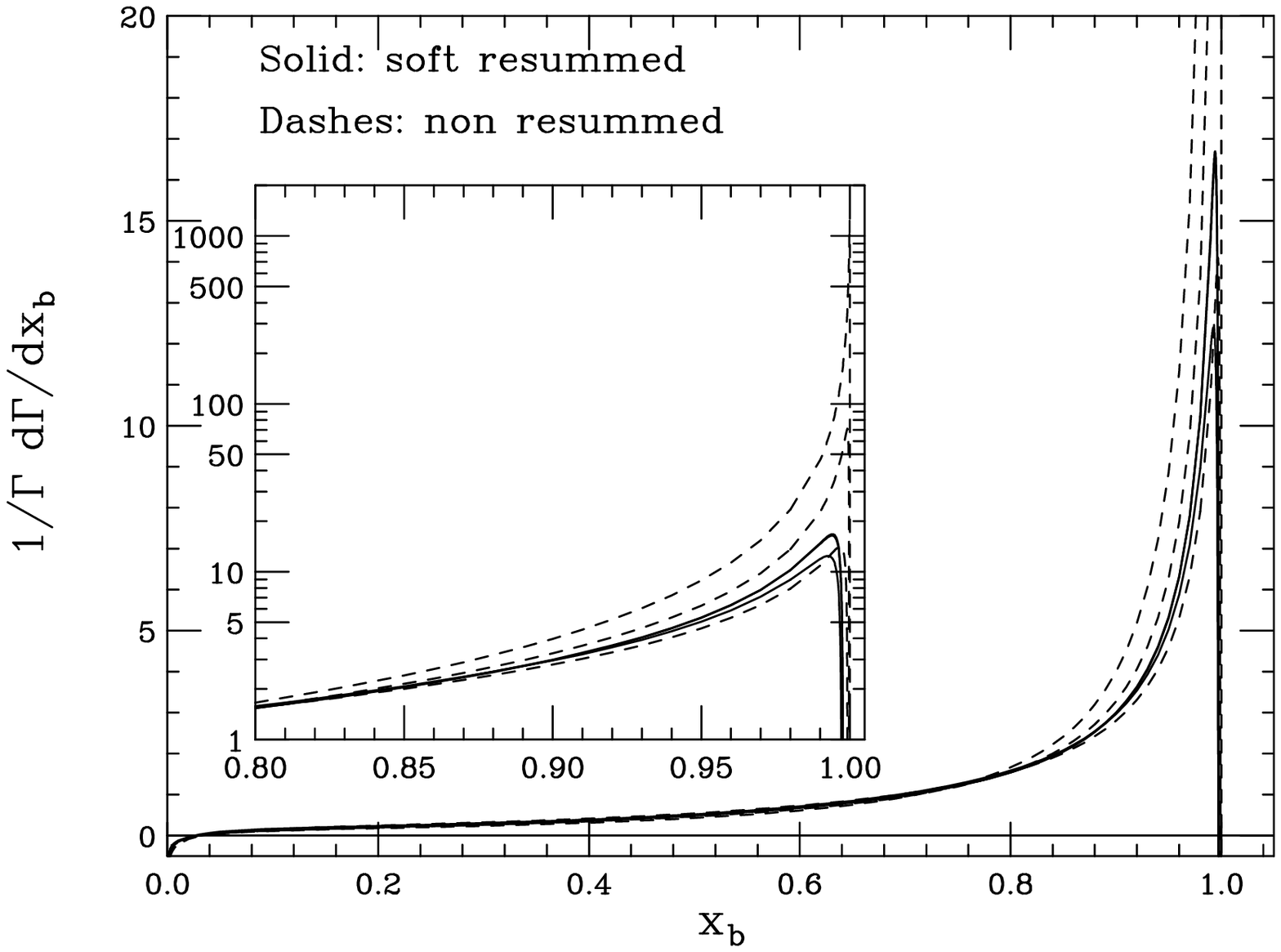,width=11.cm} 
\caption{\small As in figure~\ref{figfac}, but for different values of
$\mu_{0F}$.  The other scales are fixed
at $\mu=\mu_F=m_t$, $\mu_0=m_b$.}
\label{figfac0}}
\FIGURE[ht]{\epsfig{file=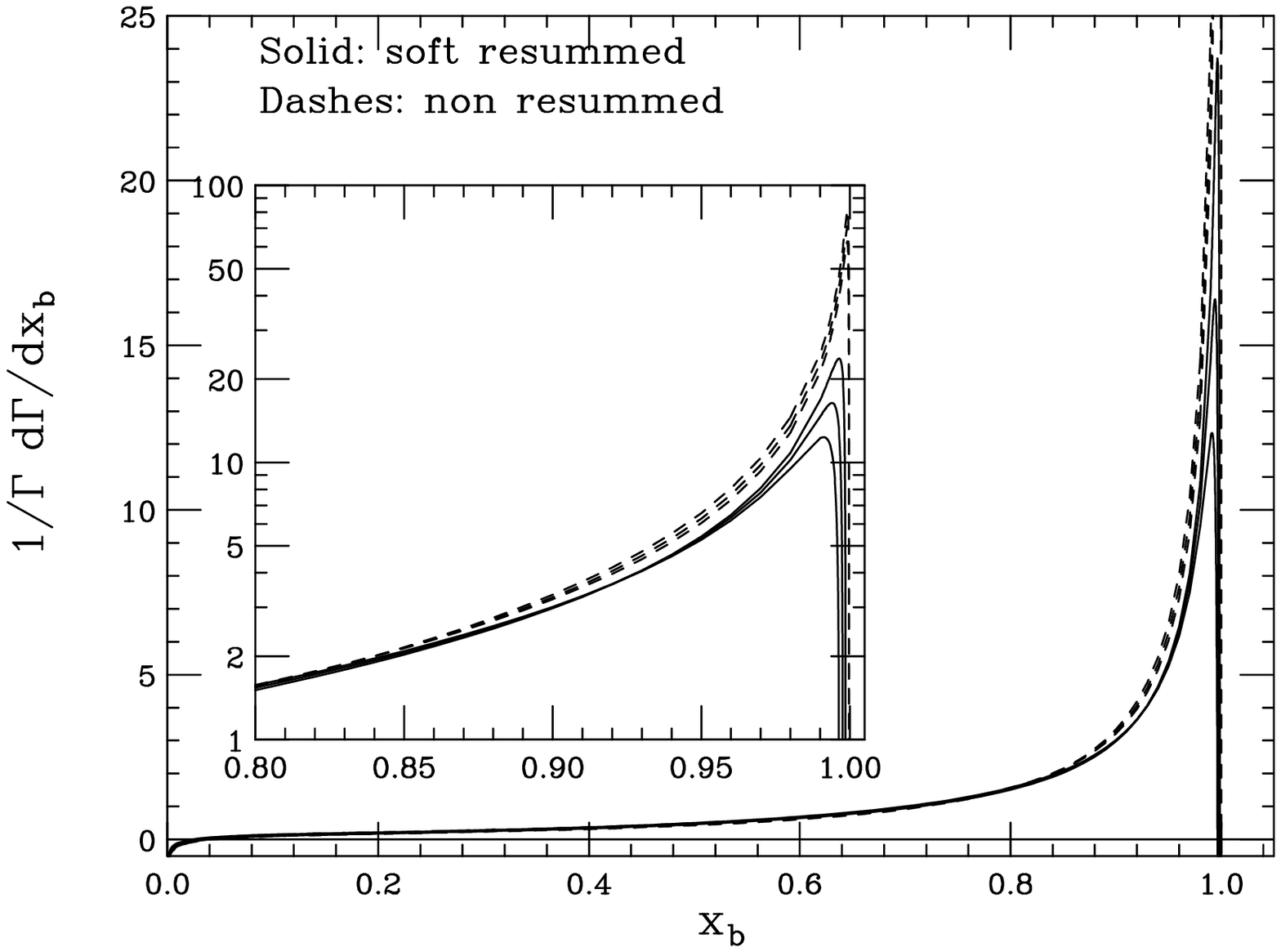,width=11.cm} 
\caption{As in figure~\ref{figfac}, but for different
values of the renormalization scale $\mu$.
The other scales are fixed at $\mu_F=m_t$, $\mu_0=\mu_{0F}=m_b$.}
\label{figren}}
\FIGURE[ht]{\epsfig{file=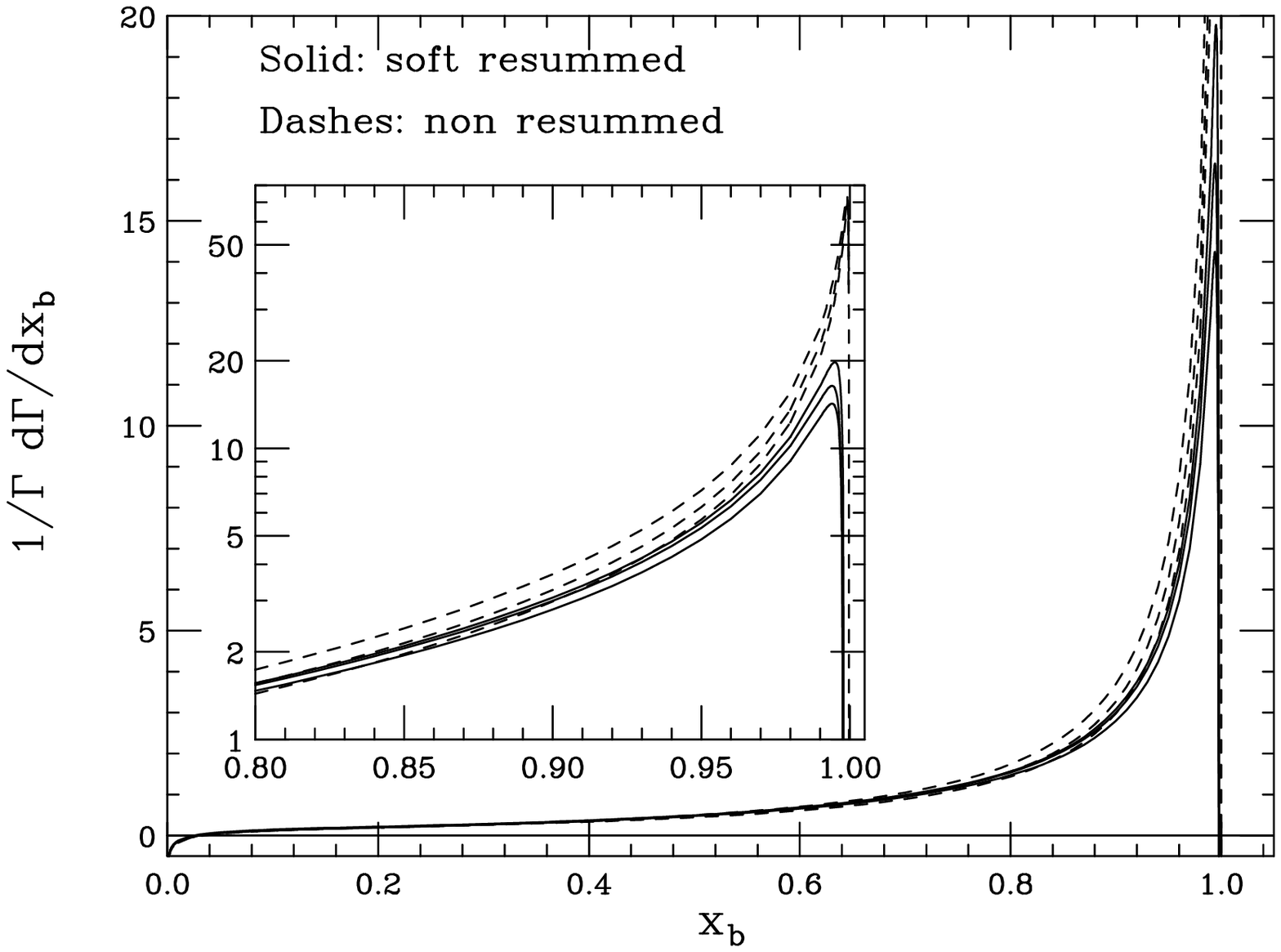,width=11.cm} 
\caption{\small As in figure~\ref{figfac}, but for different
values of the renormalization scale $\mu_0$.
The other scales are fixed at $\mu=\mu_F=m_t$, $\mu_{0F}=m_b$.}
\label{figren0}}
It is interesting to investigate the dependence of phenomenological
distributions on the renormalization and factorization scales which enter the
coefficient function ($\mu$ and $\mu_F$) and the initial condition of
the perturbative fragmentation function ($\mu_0$ and $\mu_{0F}$).
In particular, it is worth comparing the $b$-energy spectra
with and without soft resummation.
For the scales $\mu$ and $\mu_F$ we consider the
values $m_t/2$, $m_t$ and $2m_t$; for
$\mu_{0F}$ and $\mu_0$ the choices are $m_b/2$, $m_b$ and $2m_b$.
Figures~\ref{figfac} and \ref{figfac0} show the dependence of the
$x_b$ spectrum on the factorization scales $\mu_F$ and $\mu_{0F}$;
the dependence on the renormalization scales $\mu$ and $\mu_0$ is
exhibited in figures~\ref{figren} and \ref{figren0}.

We note that all distributions which include soft-gluon resummation
exhibit a reduced dependence on the factorization  and
renormalization scales.

Figure~\ref{figfac} shows that curves obtained using
different values of $\mu_F$ are almost indistinguishable once
soft resummation is included; the unresummed plots exhibit instead
a stronger effect of the chosen value for $\mu_F$.
A similar result also holds for the scale $\mu_{0F}$: the dependence
of the plots on its actual value
for $x_b>0.8$ is small if soft logarithms are resummed and
quite strong if the prediction is unresummed (figure~\ref{figfac0}).

The choice of the value for the renormalization scale $\mu$
appearing in eq.~(\ref{largen}) affects only
the neighbourhood of the Sudakov peak of the resummed predictions, at
$x_b$-values very close to one (figure~\ref{figren}), where,
as we have pointed out, our perturbative approach is anyway unreliable.
The effect of the choice of the renormalization scale
$\mu_0$ on the soft-resummed spectra
is slightly larger than the one of $\mu$
and visible at $x_b<1$ as well (figure~\ref{figren0}).
As for the non-soft-resummed predictions, although all dashed curves
in figures~\ref{figren} and \ref{figren0}
seem to converge to same point for $x_b\to 1$, the overall
dependence on $\mu$ and $\mu_0$ for $x_b<1$ is stronger than for the
resummed predictions.

As a whole, one can say that the implementation of NLL soft-gluon resummation,
along with the NLL DGLAP evolution for the perturbative fragmentation
function, yields a remarkable improvement of our phenomenological results,
since the reduced dependence on the choice of factorization and
renormalization scales in the region where the perturbative approach is
reliable corresponds to a reduction of the theoretical uncertainty.

\section{Energy spectrum of $b$-flavoured hadrons in top decay}
In this section we consider the problem of including a non-perturbative
component on top of the perturbative result, so as to make predictions
for observable
$b$-flavoured hadrons (like $B$ mesons) in top
decay. At the same time, we also
account for the inclusion of NLL soft and collinear  resummation.

We write the normalized rate for the production of $b$-hadrons
$B$ as a convolution
of the rate for the production of $b$ quarks in top decay, given by
eq.~(\ref{gamman}), and
a non perturbative fragmentation function $D^{np}(x)$:
\begin{equation}
{1\over {\Gamma}} {{d\Gamma^B}\over{dx_B}} (x_B,m_t,m_W,m_b)={1\over{\Gamma}}
\int_{x_B}^1 {{{dz}\over z}{{d\Gamma^b}\over {dz}}(z,m_t,m_W,m_b)
D^{np}\left({x_B\over z}\right)},
\label{npff}
\end{equation}
where $x_B$ is the $B$ normalized energy fraction:
\begin{equation}
x_B={1\over{1-w}}{{2p_B\cdot p_t}\over {m_t^2}},
\end{equation}
$p_B$ being the $B$ four-momentum.
Since $D^{np}(x)$ contains non-perturbative information, it cannot - for
the time being - be
calculated from first principles in QCD, but can only be
extracted from data.
We shall assume a universality property for such a function, and extract
it from fits to $B$-production data collected at LEP in $e^+e^-$
collisions. In particular, we can choose different functional forms for
$D^{np}(x)$, and  tune these hadronization
models to the data. We shall consider
a power law with two tunable parameters:
\begin{equation}
D^{np}(x;\alpha,\beta)={1\over{B(\beta +1,\alpha +1)}}(1-x)^\alpha x^\beta,
\label{ab}
\end{equation}
the model of Kartvelishvili et al. \cite{kart}\footnote
{We correct a typing mistake of ref.~\cite{cormit}, where
the normalization factor is the inverse of the correct one.
The numerical results
of ref.~\cite{cormit} are nonetheless obtained using the correct
normalization of the Kartvelishvili non-perturbative fragmentation
function.}:
\begin{equation}
D^{np}(x;\delta)=(1+\delta)(2+\delta) (1-x) x^\delta\; ,
\label{kk}
\end{equation}
and the Peterson et al. model \cite{peterson}:
\begin{equation}
D^{np}(x;\epsilon)={A\over {x[1-1/x-\epsilon/(1-x)]^2}}.
\label{peter}
\end{equation}
In eq.~(\ref{ab}), $B(x,y)$ is the Euler Beta function; in
(\ref{peter}) $A$ is a normalization constant.

In order for our procedure to be self-consistent, care must be taken to
employ the same underlying perturbative description in both the
$e^+e^-\to b\bar b$
process (where the non-perturbative contribution is fitted) and the
$t\to bW$
one (where it is used). This will be ensured by using in both
cases NLO, $\overline{\mathrm{MS}}$ coefficient functions,
along with a fully NLL soft-gluon resummed description,
with the large collinear
logarithms resummed to NLL accuracy by DGLAP evolution.
For the coefficient functions in $e^+e^-$ annihilation we shall refer to
\cite{msbar}.

Fits to data points can be performed either in $x_B$-space, or, as
recently advocated \cite{canas}, in the conjugated moment space.
When fitting in $x_B$ space
we discard data points close to
$x_B=0$ and $x_B=1$ and consider ALEPH~\cite{aleph} data\footnote{Good-quality
data are also available from SLD~\cite{sld} on $b$-flavoured mesons
and baryons.
While this paper
was being finalized, the OPAL~\cite{opal} and DELPHI~\cite{delphi}
Collaborations also published new results, which
are fully compatible with the ones from ALEPH.} in the range
$0.18\lsim x_B\lsim 0.94$.
The results of our fits are shown in table~\ref{table1}.
\TABULAR
{||c|c||}{\hline
$\alpha$&$0.51\pm 0.15$  \\
\hline
$\beta$&$13.35\pm 1.46$  \\
 \hline
$\chi^2(\alpha,\beta)$/dof&2.56/14 \\
\hline
$\delta$&$17.76\pm 0.62$ \\
\hline
$\chi^2(\delta)$/dof&10.54/15  \\
\hline
$\epsilon$&$(1.77\pm 0.16)\times 10^{-3}$\\
\hline
$\chi^2(\epsilon)$/dof&29.83/15\\
 \hline}
{\label{table1}\small
Results of fits to $e^+e^-\to b\bar b$ ALEPH data, using matched
coefficient function and initial condition, with
NLL DGLAP evolution and NLL soft-gluon
resummation.
We set $\Lambda^{(5)}=200$~MeV, $\mu_{0F}=\mu_0=m_b=5$~GeV and
$\mu_F=\mu=\sqrt{s}=91.2$~GeV. $\alpha$ and $\beta$ are the parameters in
the power law (\ref{ab}), $\delta$ refers to (\ref{kk}), $\epsilon$
to (\ref{peter}). The fits have been performed neglecting the
correlations between the data points.}

The best-fit values for the parameters of the hadronization models
are quite different from the ones quoted in \cite{cormit},
where soft-gluon resummation in the $e^+e^-$ coefficient function has not been
used in the fits.
The models in Eqs.~(\ref{ab}) and (\ref{kk}) yield very good fits to
the data,\footnote{We have also fitted the SLD data and found qualitatively
similar results. However, the value for the parameters which best fit the
SLD data are different from the ones obtained for ALEPH and quoted in
table~\ref{table1}.}
as already found in \cite{cormit}, while the model
(\ref{peter}) is marginally consistent.

Using the results in table~\ref{table1} we can give predictions for
the spectrum of $b$-flavoured hadrons in top decay. As in
ref.~\cite{cormit}, to account for the errors on the best-fit parameters,
we shall plot bands which correspond to predictions at one-standard-deviation
confidence level.
In figure~\ref{had1} we show our predictions for the $x_B$ distribution using
the three models fitted to the ALEPH data.
At one-standard-deviation confidence level, the three predictions are
different,
with the Peterson model yielding a distribution which lies quite far from
the other two and peaked at larger $x_B$-values.
Within two standard deviations, the predictions obtained using the
models (\ref{ab}) and (\ref{kk}) are nonetheless in agreement\footnote{The
differences between the various models mainly originate from the
varying quality of the fits to $e^+e^-$ data where, as the $\chi^2$ values
in table~\ref{table1} seem to suggest, a given model is sometimes
not really able to describe
the data properly, due to its too restrictive functional form.}.
We also note that figure~\ref{had1} is qualitatively similar to figure~4 of
ref.~\cite{cormit}, where soft-gluon resummation in the coefficient function
had not been implemented. 
In fact, the different perturbative content of these curves is now
compensated by different values for the non-perturbative parameters
$\alpha$, $\beta$, $\delta$ and
$\epsilon$, accordingly set by the fitting procedure. This hadron-level
similarity does however not lessen the importance of the higher degree of
reliability of the perturbative-level calculation provided for by the
soft-gluon resummation: Such an accurate calculation can be used as a
firmer starting point for testing and fitting non-perturbative models.
On the other hand, the inclusion of 
perturbative resummation cannot be expected to improve 
the agreement between hadron-level results obtained with
different phenomenological non-perturbative ans\"etze.

\FIGURE
{\epsfig{file=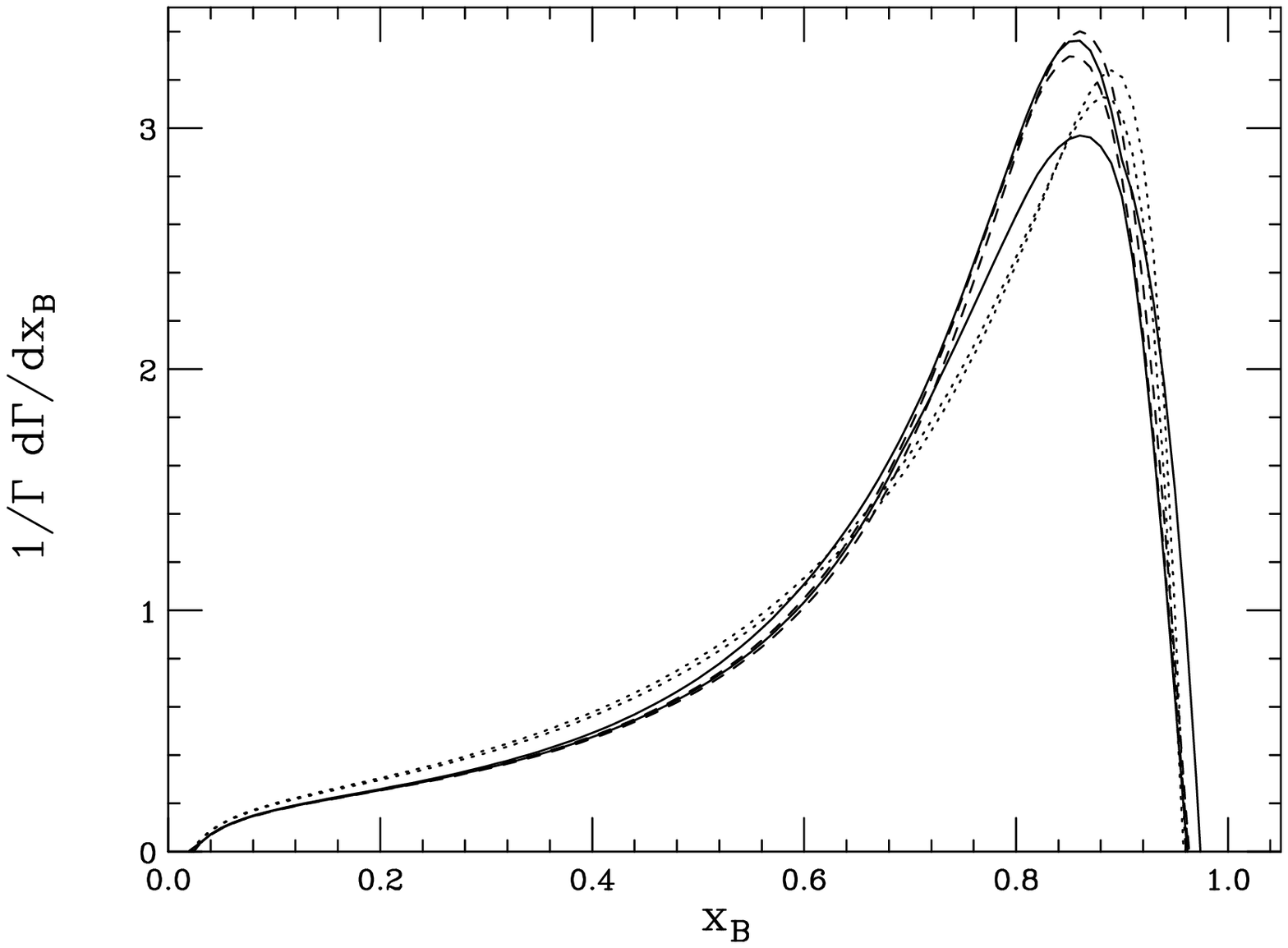,width=11.cm} 
\caption{$x_B$ spectrum in top decay, with the hadronization
modeled according to a power law (solid lines),
the Kartvelishvili et al. (dashes) and the Peterson (dots) model,
with the relevant parameters fitted to the ALEPH
data. The plotted curves are the edges of bands at
one-standard-deviation confidence level. NLL soft-gluon resummation
is included.
We set $\mu_F=\mu=m_t$ and $\mu_{0F}=\mu_0=m_b$.}
\label{had1}}

An alternative, and probably better, way of determining and
including non-perturbative
information makes use of moment
space perturbative predictions and data \cite{canas}.
The full hadron-level result
can be written in $N$-space as the product of a perturbative and a
non-perturbative contribution, $\Gamma^B_N = \Gamma^b_N D^{np}_N$. For
each value of $N$ one can then extract the corresponding $D^{np}_N$
value from $e^+e^-$ data, with no reference whatsoever to a specific
hadronization model, and use it to predict the same
moment in top decay. The DELPHI Collaboration~\cite{delphi} has
recently published preliminary results for the moments of $B$-meson
fragmentation in $e^+e^-$ collisions up to $N=5$. From these data, and
using the moments of the $e^+e^-$ perturbative contributions~\cite{cc},
one can extract $D^{np}_N$. The corresponding $\Gamma^B_N$ values can
then be calculated making use of the results for $\Gamma^b_N$ obtained
in this paper. Calling $\sigma^B_N$ and $\sigma^b_N$ the moments for
the production rate of
$B$ mesons (measured) and $b$ quarks (calculated in
perturbative QCD) in $e^+e^-$ annihilation, we have
$\sigma_N^B=\sigma_n^B D^{np}_N$
and hence
\begin{equation}
\Gamma^B_N = \Gamma^b_N D^{np}_N =
\Gamma^b_N \frac{\sigma^B_N}{\sigma^b_N} \;.
\end{equation}

Table~\ref{table2} shows a practical implementation of this procedure.
Predictions for the moments $\Gamma^B_N$ of $B$-meson spectra in
top decay are given, making use of the DELPHI experimental data.
Two sets of perturbative results ([A] and [B]) are shown,
the first using $\Lambda^{(5)} = 0.226$~GeV and $m_b = 4.75$~GeV, the second
using the default parameters of this paper. As expected, the
perturbative calculations and the corresponding non-perturbative
components differ, but the final predictions for the physical results 
$\Gamma^B_N$ are to a large extent identical.
\begin{small}
\TABULAR
{| c | c c c c |}
{\hline
& $\langle x\rangle$ & $\langle x^2\rangle$ & $\langle x^3\rangle$
& $\langle x^4\rangle$ \\
\hline
$e^+e^-$ data $\sigma_N^B$&0.7153$\pm$0.0052 &0.5401$\pm$0.0064 &
0.4236$\pm$0.0065 &0.3406$\pm$0.0064  \\
\hline
\hline
$e^+e^-$ NLL $\sigma_N^b$ [A]   & 0.7666 & 0.6239 & 0.5246 & 0.4502  \\
$e^+e^-$ NLL $\sigma_N^b$ [B]   & 0.7801 & 0.6436 & 0.5479 & 0.4755  \\
\hline
$D^{np}_N$ [A]          & 0.9331 & 0.8657 & 0.8075 & 0.7566 \\
$D^{np}_N$ [B]          & 0.9169 & 0.8392 & 0.7731 & 0.7163 \\
\hline
\hline
$t$-decay NLL $\Gamma^b_N$ [A]& 0.7750 & 0.6417 & 0.5498 & 0.4807 \\
$t$-decay NLL $\Gamma^b_N$ [B]& 0.7884 & 0.6617 & 0.5737 & 0.5072 \\
\hline
$t$-decay $\Gamma^B_N$ [A]        & 0.7231 & 0.5555 & 0.4440 & 0.3637 \\
$t$-decay $\Gamma^B_N$ [B]        & 0.7228 & 0.5553 & 0.4435 & 0.3633 \\
\hline}
{\small\label{table2}Experimental data for the moments
$\sigma^B_N$ from
DELPHI~\protect\cite{delphi}, the resummed $e^+e^-$ perturbative
calculations for $\sigma^b_N$~\protect\cite{cc},
the extracted non-perturbative contribution
$D^{np}_N$. Using the perturbative results $\Gamma^b_N$, a prediction for
the physical observable moments $\Gamma^B_N$ is given.
Set [A]: $\Lambda^{(5)} = 0.226$~GeV and $m_b = 4.75$~GeV,
set [B]: $\Lambda^{(5)} = 0.2$~GeV and $m_b = 5$~GeV. The experimental
error should of course be propagated to the final prediction.}
\end{small}
\section{Conclusions}
We have discussed soft-gluon resummation in top quark decay within the
framework of perturbative fragmentation which, by making use of
the DGLAP evolution equations, allows one to resum with NLL accuracy the
large logarithms $\alpha_S\ln(m_t^2/m_b^2)$ which appear in the
${\cal O} (\alpha_S)$ massive calculation.
The $\overline{\mathrm{MS}}$ coefficient function and the initial condition
of the perturbative fragmentation function contain terms which become
arbitrarily large for soft-gluon radiation. Such soft terms are process
dependent in the coefficient function and process independent in the
initial condition of the perturbative fragmentation function.

We have performed the resummation of soft-gluon effects
in the coefficient function of the top-to-massless-quark decay process
with NLL accuracy,
and matched the resummed result
to the full ${\cal O}(\alpha_S)$ one. This result has then been
combined with the one of  ref.~\cite{cc}, which resums NLL
process-independent soft-gluon
contributions in the initial condition of the perturbative
fragmentation function, to produce a resummed prediction for the full
top-to-massive-bottom decay.

We have presented the resummed $b$-quark energy spectrum in top-quark
decay and compared it with the unresummed prediction. We have found
that at large $x_b$ the implementation of soft-gluon resummation has a
visible impact: The $x_b$ spectrum is smoothed out and shows the
characteristic Sudakov peak. Our prediction for the $b$-energy spectrum
is negative for $x_b\to 0$ and $x_b\to 1$; we have interpreted the
behaviour at small $x_b$ as due to unresummed $\ln x_b$ contributions
in the coefficient function and at large $x_b$ to missing
non-perturbative effects.

We have investigated how the $b$-energy spectrum varies if we choose
different values for the factorization and renormalization scales which
enter our calculation. We have found that after including NLL
soft-gluon resummation the distributions exhibit very little dependence
on the choice of such scales, which corresponds to a reduction of the
theoretical uncertainty of our prediction.

Finally, we have made use of $b$-flavoured hadron data in $e^+e^-$
collisions to extract a non-perturbative contribution, and we used it to
calculate hadron-level predictions in the top decay process. This has
been done both in $x_B$ space, tuning various hadronization models to
$e^+e^-$ distributions, and in $N$ space, directly extracting the
values of a few moments of the non-perturbative fragmentation function.
In both cases care has been taken to make the procedure self-consistent
by employing the same perturbative description in both the $e^+e^-$ and
top decay processes.

\acknowledgments
We are especially grateful to S. Catani for many useful conversations
on the topic of soft-gluon resummation. We also acknowledge discussions
with U. Aglietti, L.H. Orr, M.H. Seymour and B.F.L. Ward on
these and related topics.

\end{document}